\documentclass[journal, 10pt]{IEEEtran}

\usepackage{graphicx} 
\usepackage{biblatex}
\usepackage{microtype}
\usepackage{hyperref}
\usepackage{acronym}
\usepackage{booktabs}
\usepackage{csquotes}
\usepackage{amsmath}
\usepackage{subcaption}
\usepackage{xcolor}
\usepackage[inline]{enumitem}

\acrodef{fmv}[FMV]{the Swedish Defence Matériel Administration}
\acrodef{foi}[FOI]{the Swedish Defence Research Agency}
\acrodef{mal}[MAL]{Meta Attack Language}
\acrodef{rl}[RL]{reinforcement learning}

\acrodef{wmi}[WMI]{Windows Management Instrumentation}

\acrodef{gnn}[GNN]{graph neural network}
\acrodef{mpnn}[MPNN]{message-passing neural network}
\acrodef{cnn}[CNN]{convolutional neural network}
\acrodef{mlp}[MLP]{multi-layer perceptron}
\acrodef{ids}[IDS]{intrusion detection system}
\acrodef{nids}[NIDS]{network intrusion detection system}
\acrodef{ppo}[PPO]{proximal policy optimization}
\acrodef{gae}[GAE]{generalized advantage estimation}
\acrodef{wl}[WL]{Weisfehler-Lehman}
\acrodef{mdp}[MDP]{Markov decision process}
\acrodef{ippc}[IPPC]{International Probabilistic Planning Competition}
\acrodef{pddl}[PDDL]{Planning Domain Definition Language}
\acrodef{kg}[KG]{Knowledge graph}
\acrodef{rddl}[RDDL]{Relational Dynamic Influence Diagram
Language}
\acrodef{iid}[IID]{Independent and Identically Distributed}
\acrodef{ood}[OOD]{Out of Distribution}
\acrodef{rdf}[RDF]{Resource Description Framework}
\acrodef{gui}[GUI]{Graphical User Interface}
\acrodef{ml}[ML]{machine learning}
\acrodef{pomdp}[POMDP]{partially observable Markov decision process}
\acrodef{llm}[LLM]{large language model}
\acrodef{it}[IT]{information technology}

\newacroindefinite{mdp}{an}{a}
\newacroindefinite{mpnn}{an}{a}
\newacroindefinite{mlp}{an}{a}
\newacroindefinite{rl}{a}{a}
\newacroindefinite{mal}{a}{a}

\newcommand{\crate}{Crate}
\newcommand{\lore}{Lore}
\newcommand{\network}{ADS-24}

\newcommand{\cratelang}{CadsLang}

\title{A Cyber Range Evaluation of\\ Inductive Autonomous Network Incident Response Agents}
\author{
  Jakob Nyberg\thanks{Jakob Nyberg (jaknyb@kth.se), Pontus Johnson and Mathias Ekstedt
    are with the Department of Network and Systems Engineering at KTH Royal Institute of
    Technology in Stockholm, Sweden. Andrei Buhaiu and Joakim Loxdal were with KTH at the
  time of the work.}%
  , Teodor Sommestad\thanks{Teodor Sommestad is with the Swedish Defence Research Agency FOI.}%
  , Andrei Buhaiu%
  , Joakim Loxdal%
  , Pontus Johnson%
  , Mathias Ekstedt%
}

\begin{document}







\maketitle

\begin{abstract}

  We have applied a novel combination of classic
  relational database modeling
  and modern reinforcement learning to
  automated intrusion response to
  enable inductive decision agents.
  Unlike previous work in the area,
  our approach allows
  agents to handle variable quantities
  of network assets without retraining.
  We test the operational performance of our approach in
  a virtualized
  computer network
  with variable topologies,
  red-team automation and
  simulated user agents.
  The task of the agents is to prevent hosts in the network
  from being accessed by the red-team agent,
  while minimizing the availability costs induced from defensive measures.
  We test a combination of heuristic agents
  and policies trained using relational reinforcement learning.
  The reinforcement learning agents were overall more efficient at defending the system
  than the heuristic policy,
  and the closest to the performance of human operators at the
  task.
\end{abstract}


\begin{IEEEkeywords}
  autonomous cyber defense, sim-to-real, cyber defense simulation, cyber attack
  simulation, reinforcement learning, machine learning, graph neural network, digital twin
\end{IEEEkeywords}


\section{Introduction}


Security incidents involving computer networks are an unfortunate but regular occurrence
in our digitalized and connected society~\cite{msb}.
Rapid mitigation of attacks on \ac{it} networks is important to avoid sensitive
information being stolen, or critical systems made inoperable.
Mitigation of network security incidents is traditionally done manually by network
operators, or automatically through pattern- and rule-based systems.
\Ac{ml} methods aim to learn patterns and rules automatically from system data, and a
number of efforts have already been made to apply different \ac{ml} methods for
mitigating cyber security incidents~\cite{DBLP:conf/cnsm/HammarS20,
  https://doi.org/10.1002/aaai.70021, 11593294, https://doi.org/10.1002/aaai.70021,
hicksBuildingBetterEnvironments2026, drasar24}.
A common design among control-theoretic approaches is that the defender agent should
maintain system operations while minimizing costs from security violations.
To not make the problem trivial, actions taken by the defensive agents are assumed to
incur some form of cost, typically proportional to the disruption to the system the action causes.
We have followed this general approach to the task, but have attempted to address two
issues we have identified in previous works: Evaluations done purely in simulation, and
implicit assumptions that the network is static.
\begin{figure}[tpb]
  \centering
  \includegraphics[width=\linewidth]{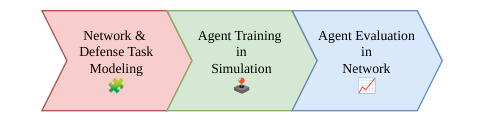}
  \caption{High-level rendering of the three steps we follow for sim-to-real development
  of automated cyber defense agents.}\label{fig:simplepipeline}
\end{figure}
\newpage
\subsection*{Problem 1: Simulated Evaluation Environments}
Several previous academic works have proposed methods for using \ac{rl} to
handle network intrusion response tasks~\cite{320855, 10.1145/3729213,
drasar24}.
However, most of these were evaluated purely in simulation,
with no consideration of more realistic operating conditions.
This places a significant amount of faith in assumptions made
by the simulation model about the dynamics of the target system,
and what data will be available to the agent where it is to be used.
To avoid the pitfalls associated with simulator-only evaluations, we test automated
defender agents with the cyber range \crate{}~\cite{10.1007/978-3-030-70852-8_12,
9229649}, where we instantiate a network of virtual machines with simulated user agents.

\subsection*{Problem 2: Static Data Models}
Though simulated evaluations are common, there have been a some works that evaluate
\ac{rl} agents in computer networks~\cite{ITURBE2026129168, 10.1145/3689933.3690834,
KYRIAKOPOULOS2026105058, HUGHES2022103150, 10.1145/3664476.3670917}.
However, all implicitly assume a fixed network topology through their state representation
format.
While this assumption may be valid for certain industrial systems, most \ac{it} networks
are dynamic with both physical and logical assets such as hosts, users or services being
added and removed over time. This leads to agents needing to be frequently retrained to
account for network changes.

We address the data modeling issues by using a relational
state representation to enable online insertion
and removal of entities, attributes and relations.
The relational states are used in combination with inductive graph
representation learning methods to avoid having to retrain policies
when the database changes.
Graph-based approaches to the intrusion
response problem have been proposed previously, but these have only
been tested in simulation~\cite{DBLP:conf/csr2/NybergJ24, 11593294} and fail to address
how the graphs should be constructed in a more realistic environment.

\newpage

\subsection*{Summary of Approach}

We implement relational \ac{rl} agents using Vejde, a reinforcement learning framework
for decision problems with relational state spaces~\cite{nyberg2026vejde}.
The agents were developed using a sim-to-real procedure, as illustrated
in~\autoref{fig:simplepipeline},
where we trained agents in simulation
before evaluating them with
the computer network environment.
As our simulation environment, we have used the \emph{\ac{mal}
Simulator}~\cite{loxdal_2026_hp777-8rw38}, for which
we created models of \network{}
using \ac{mal}~\cite{DBLP:conf/IEEEares/JohnsonLE18}.

\begin{figure*}[htpb]
  \centering
  \includegraphics[width=0.8\textwidth]{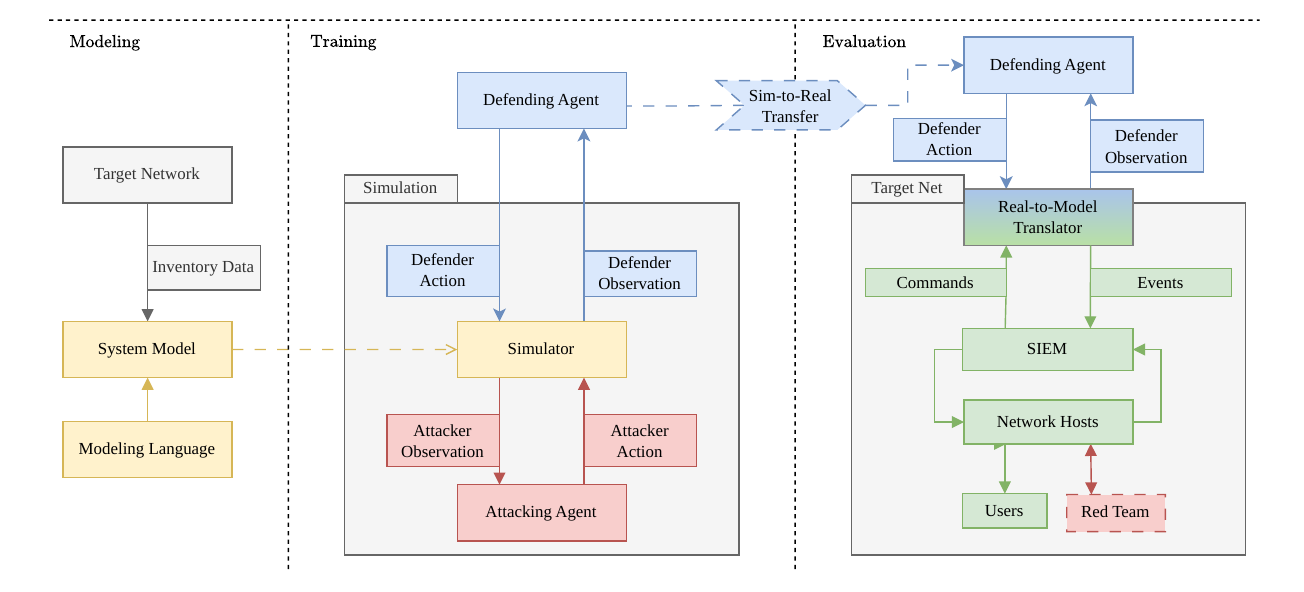}
  \caption{Conceptual flow chart over the \emph{sim-to-real} defender agent development
    and evaluation process described in this work.
    First, a data model describing a network is created with a given modeling language
    and collected inventory data.
    The model is then used with a simulator, which may use simulated attackers, to
    optimize a defending agent policy.
    Finally, the agent is evaluated with the target network, containing both adversarial
    and neutral agents.
    A translation interface converts and classifies events from the network into the data
    model used by the defending agent,
    and translates the agent's actions back to commands to be executed on network
  hosts.}\label{fig:pipeline}
\end{figure*}

Our computer network, named ADS-24, consisted of 37 virtual machines.
We put the network under attack for two hours using the red-team automation tool
\lore{}~\cite{9740000},
which the defender agent should stop
by shutting down and isolating hosts in the network.
The network also contains a number of simulated user agents
that generate additional activity by perform various work-related tasks.
A more detailed flow chart illustrating our implementation is shown in \autoref{fig:pipeline}.

We compare the relational \ac{rl} agents with a manually crafted heuristic agent that
uses the same type of relational database as input.
Five human subjects with prior experience in log analysis
were also assigned the defense task.
Our results show that one of the \ac{rl} agents consistently outperforms the heuristic
agent in regard to minimizing the joint cost, and prevents \lore{} from completing its attack path.
The heuristic policy is effective at blocking \lore{}, but incurs a high availability
cost since it indiscriminately acts against all observed alerts. However, all agents are
worse at handling false positives than the human operators, who managed to produce
the lowest overall costs.
From our results, we conclude that there is potential in using \ac{rl} for automated
defensive policies, but that there is still a significant amount of work remaining for
the approach to be of practical use.
Our experiments have also produced a number of research artifacts that we have made
public. These include:
\begin{itemize}
  \item The \ac{mal} language~\cite{buhaiu_2026_rpcde-z6r38} and network system model.
  \item A Vejde/MAL Simulator interoperability library~\cite{nyberg_2026_gq7cc-2ww92}.
  \item An interface to map network events to \ac{mal} attack steps~\cite{nyberg_2026_enqp2-38375}.
  \item Data collected from the network environment~\cite{nyberg_2026_4mh6z-zw065}.
\end{itemize}

\section{Related Work}


\subsection{Similar Evaluations}

As mentioned in previously, most related works in the area of \ac{rl} for intrusion response
perform evaluations with simulated environments. We have, however found a handful of
works that both train and evaluate agents in more realistic environments, which we cover
in this section. The task definitions of all these works are similar, with a network being 
attacked and a defending agent minimizing the cost of defense.

\textcite{ITURBE2026129168} trained and evaluated agents
with a network of virtual machines orchestrated by Proxmox. 
Their network consisted of four hosts and a digital measurement system. 
MITRE Caldera was used to automate a cyber attack on the network.
The observation vector provided to the defender agents 
was defined as a concatenation of a numeric security state 
for each asset in the network.
There were no users in the network apart from the attacker
and there is no mention of false positive alerts in the text.

The authors define 13 action types which can
be applied to different assets. These include disabling
hosts, signing out user and deleting files.
Defensive agents were trained with \ac{ppo} and Deep Q-learning using transitions gathered online from the network environment. 
The observation and action spaces were ordered 
and of fixed lengths, meaning that the approach 
functionally assumes the network topology to be static.
This also means that the agents would need to be
retrained if the network topology changes.
It is unclear how long training took in practice, 
though the authors mention that the time was not
insignificant.

\textcite{10.1145/3689933.3690834} and the subsequent work by
\textcite{KYRIAKOPOULOS2026105058} also used Caldera for 
attacker automation, but with the software 
GNS3 to emulate a computer network of five hosts.  
They model an attack path and defenses as a MulVal attack
graph. Agent observations consisted of a vector indicating
which of attack graph nodes have been performed by the attacker, 
and which defenses have been enabled.
Their approach thus also suffer 
from the same static data assumptions as the one by~\textcite{ITURBE2026129168}.
Training was done with tabular Q-learning and \ac{ppo}.

\textcite{HUGHES2022103150} used GNS3 
for training and evaluation, but with a
custom scripted approach to automate attacks.
Their network consisted of four hosts 
and implemented seven defensive action types, mainly related to
disabling hosts and network interfaces.
\textcite{10.1145/3664476.3670917} uses an almost identical approach, but with a different network topology, attack chain and action types.
Both works define the state as an aggregation over
network asset attributes, 
such as the sum of generated alerts and nodes that were online. 
This avoids some of the data modeling
issues in the previously described works, 
and reduces the overall number of states. 
However, their action selection mechanism
still assumes a static network, 
and the simple aggregation of the network attributes
leads to all locality of 
network events being lost.

In summary, out of the works that we have found which evaluate
\ac{rl} agents for intrusion response in computer networks, all
have data modeling issues that impact the practicality of the system. 
This does not invalidate the entire approach, but 
rather that the authors have overlooked 
important components of data modeling and machine learning. 
In most cases, a relational or factored input 
representation format, such as the one we employ, 
could be used as part of their respective approaches
to address these issues.
We also want to note that all works
use computer networks with less than ten hosts. 
This puts the scalability of the approaches into question,
especially regarding the time required to train the agents.

\subsection{Cyber Defense Simulators}

Due to the amount of time required to train 
\ac{rl} agents with virtualized network environments, 
we have opted to use a simulator to train agents 
before deployment in the target system.
This is sometimes referred to as a \emph{sim-to-real} approach in 
literature~\cite{DBLP:journals/jair/KirkZGR23,
  DBLP:conf/nips/WagenmakerHKJ024}, 
  where agents are trained in a simulator of the target system, with the intent
of zero-shot, or few-shot, transfer of the agent into the target
system. Another reason for using a simulator is that we only implemented one network, 
and we wanted to use the one we had to test the agent's capability to generalize.

Within the research area of
automated cyber defense, 
several simulation environments intended for
training \ac{rl} agents have been proposed. These include
CybORG~\cite{https://doi.org/10.1002/aaai.70021, emerson2024cyborg},
Yawning Titan~\cite{yawning} and
CyberWheel~\cite{DBLP:conf/uss/OeschCWDSRWA24}, among others~\textcite{drasar24, 10.1145/3729213}.
However, most works we have found using these simulators only evaluate their approach using the simulator, often due to the 
abstract nature of the simulator making the transfer difficult~\cite{hicksBuildingBetterEnvironments2026}.
It is also common to do what~\textcite{DBLP:journals/jair/KirkZGR23} 
call \emph{singleton} testing, meaning that the problem used for training is also the
problem tested on. Singleton testing is useful to confirm that the
policy is optimal, but it does not test how the policy responds to
changes in the problem. 
%


\section{Preliminaries}

\subsection{The Meta Attack Language}\label{sec:mal}

We use the \acf{mal}~\cite{DBLP:conf/IEEEares/JohnsonLE18, WIDEL2023103284} to define a
data model of the problem domain. \ac{mal} is a modeling language with the intended use
of creating domain-specific languages to represent networked systems and techniques used
in cyber attacks.
There exists a number of \ac{mal} languages for various domain applications, such as
CoreLang~\cite{KATSIKEAS2024104057} to model generic network environments.
A \ac{mal} language defines a set of abstract \emph{asset} types to describe elements of
a system, and which can define relations to other assets.
Asset types may also be associated with \emph{attack steps}, representing actions and
techniques that the asset can be the subject of.
For instance, CoreLang defines an \enquote{eavesdrop} attack step for the asset type
\enquote{Network} to represent the act of sniffing traffic.
Attack steps can have causal relationships with other attack steps, meaning that one
attack step being performed suggests that subsequent attack steps, either on the same or
related assets, are possible to perform.
As an example, the aforementioned \enquote{eavesdrop} step is a parent to the
\enquote{attemptRead} step of any \enquote{Data} asset associated with the \enquote{Network} asset.
\Iac{mal} language can also define a set of \emph{defense steps}, which represent
defensive techniques to block attack steps from being executed.
Given a \ac{mal} language, the asset classes can be instantiated to define an
\emph{instance model}; a model of a particular system in terms of asset types defined in
the language.
The instance model can be used to generate an \emph{attack graph}, which contains all
attack step paths that are possible to perform given the assets and relations in the
instance model, as well as the attack step relations defined in the language.

\subsection{Markov Decision Processes}\label{sec:mdp}


Following previous works in the domain of autonomous cyber
defense~\cite{10.1145/3689933.3690834, https://doi.org/10.1002/aaai.70021,
DBLP:conf/cnsm/HammarS20}, we treat the defense task as an episodic, relational \acf{pomdp}.
A \acf{mdp} is a formal model of a sequential decision problem~\cite{DBLP:books/wi/Puterman94}.
The process is divided into discrete \emph{states}, in which sets of \emph{actions} can
be taken by a decision-taking \emph{agent}.
A probability distribution models the likelihood of reaching different states, given a
state and an action by the agent.
A \emph{cost}, or reward, value is used to measure the agent's success at the problem.
An optimal policy is a strategy of selecting actions given the state that minimizes the
expected, optionally discounted, cost over time.
\Iac{pomdp} extends the \ac{mdp} definition to model decision problems where the state is
not directly observable, and defines a set of \emph{observations} and a probability
distribution over observations,
which may be influenced by the hidden state or actions from the agent.

\Iac{mdp} tends to model a singular decision problem or
system~\cite{DBLP:books/wi/Puterman94, DBLP:journals/jair/KirkZGR23}.
However, as we assume that the system may change
across time and different environments,
such as with differing network constellations and adversary behaviors.
We therefore believe it is more useful to formulate the problem as a \emph{distribution}
or collection of multiple \acp{mdp}.
Various attempts have been made at extending the \ac{mdp} formulation to facilitate
this~\cite{DBLP:journals/jair/KirkZGR23}.
One such definition is the \emph{relational} MDP~\cite{DBLP:series/faia/2009-192},
which is \iac{mdp} defined
in terms of \emph{lifted} first-order logic. This allows us to specify a model of the
problem without needing to declare the number
of assets in the system, which may be unknown to us.
A relational MDP may instantiate several \emph{grounded} \acp{mdp} by substituting the
lifted variables with concrete values and object identifiers.
An optimal policy for a relational \ac{mdp} will thus be optimal for a collection of
problems, but not necessarily on each respective grounding, as there may exist an optimal
grounded policy for that particular problem instance.
A policy based on a concatenation of factors, as seen in related works,
can be considered a grounded policy.

A related definition to the relational \ac{mdp} is the \emph{factored} \ac{mdp}.
A factored \ac{mdp} assumes the state to be composed of a set of discrete state
variables, and that the value function can be decomposed into a linear combination of
basis functions.
The functions may individually only depend on a subset of variables from the full
state~\cite{DBLP:journals/jair/GuestrinKPV03},
meaning that the problem of learning the value function can be
separated into smaller semi-independent problems.
We make use of both definitions and define the \ac{mdp} state as
a set of discrete entities, defined by a threat modeling language,
which are sparsely related through logical relations derived from
network event data. This format can also be perceived as a relational database, with a given schema.

\subsection{Vejde}\label{sec:vejde}

Using a relational state representation provides a number of data modeling benefits,
but also induces new challenges in how the data should be handled with \ac{rl} methods.
Common \ac{rl} frameworks, such as Stable Baselines 3 and RLLib do not support
graph-based or dynamically sized observations, for instance.

We use Vejde, a reinforcement learning library designed for decision problems with
factored and relational states~\cite{nyberg2026vejde}.
In Vejde, states and observations are represented as sets of facts expressed in
first-order logic, which are converted to bipartite factor graphs for message passing.
Within the context of this work, predicates and entity types in the state database are
expressed in \ac{mal}.
Entities represented in the bipartite graph are encoded to
latent representations using message passing neural networks,
and a policy function computes action probabilities from the
entity representations.
The sizes of the action and observation spaces are determined
by the data model used to describe the problem domain,
rather than a particular problem instance. This allows us to avoid the use of fixed-sized
input and output common in similar works~\cite{HUGHES2022103150, KYRIAKOPOULOS2026105058,
ITURBE2026129168}.
To facilitate inductive policies,
only the type classes of entities
are observed by agents,
and not specific identifiers.
This means that two hosts with different identifiers \enquote{ap01} and \enquote{ap02}
will be represented as identical \enquote{Application} assets in the state.
They may, however, be differentiated by
their relations, such as when an attack step
is observed on one of the hosts.

\subsection{The Cyber Range Crate}\label{sec:crate}

Crate is a computer network emulation platform built and maintained by
\ac{foi}~\cite{9229649}.\ \crate{} instantiates networked virtual machines based on a
description language through a combination of proprietary scripts and open-source tools.
The instantiation process includes scripts for managing Windows domain settings, creating
users based on name lists, software installation via Chocolatey, and system configuration
using Ansible.

The result of this process is one or more computer networks, typically used for exercises
in cyber security.
Since 2008, the platform has been used in several technical tests, national exercises,
international exercises, and battle readiness exercises.
Recent examples of large scale cyber defence exercises include the collaborative cyber
defense exercise Safe Cyber,
and performance assessments for the Swedish Armed Forces~\cite{holm2025realistic}.  Crate
manages network states through virtual machine snapshots.
A snapshot can be taken of an instantiated computer network after it has been deployed,
or after it has been manually adjusted.
The deployed network can then be restored to the snapshot state at any time.

Various tools are implemented in \crate{} to orchestrate events in the event plane.
This includes the red-team emulation tool \lore{}, which uses a combination of
machine-learning models and heuristics to automatically attack computer networks~\cite{9740000}.
Lore's behavior can be configured through a threat agent profile and a scenario configuration.
The threat agent profile allows adjustment of general behaviors, such increasing the
tendency to scan networks.
The scenario configuration makes it possible to adjust priorities based on specific
network information, such as focusing on specific IP addresses.
Legitimate user events are simulated by procedurally executing commands and interacting
with graphical components within host user sessions.

\section{Network \& Task Description}\label{sec:network}

\begin{figure}[tpb]
  \centering
  \includegraphics[width=0.48\textwidth]{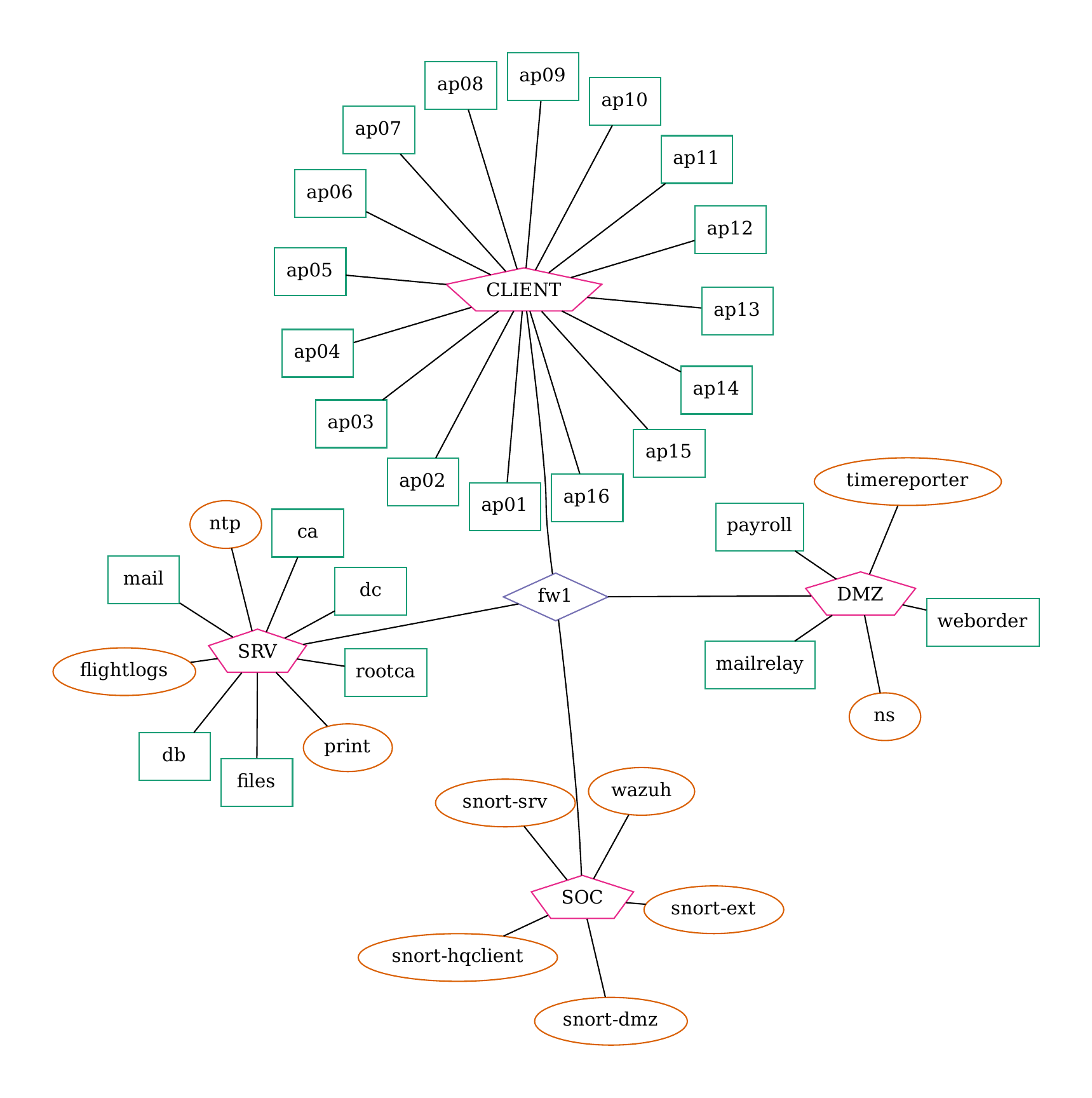}
  \caption{Topology of the AIR-DELIVERY-SYSTEM24 (ADS-24) computer network represented as a graph.
    Windows hosts are denoted with green rectangles, and Linux host with orange
  ovals.}\label{fig:network-map}
\end{figure}

The network we have implemented using \crate{} was named \enquote{AIR-DELIVERY-SYSTEM24}
(\network{}).
The intended functions of \network{} are to keep track of maintenance needs, purchase new
spare parts and store costs and salaries.
ADS-24 consists of four subnets, identified as \enquote{CLIENT}, \enquote{DMZ},
\enquote{SRV} and \enquote{SOC}, which are all connected through a shared firewall server.
The 37 machines across the subnets run either Linux or Windows, with machines running
Windows being the most frequent.
The SOC network contains machines related to monitoring and is inaccessible to all agents
to not disturb logging during experiments.
The layout of \network{} is illustrated as a graph in \autoref{fig:network-map}.

The defender agent should protect \network{} against an ongoing intrusion by an
adversarial entity for a fixed period of time, while minimizing the operational costs
caused by the attack and the defense. The time period was set to two hours, as this was
approximately the time required for our red team automation system to reach its goals
when no defenses were present.

\subsection{Alert Rules
\& Active Responses}

Wazuh runs in \network{} for event logging and issuing commands to hosts.
Each host, except \enquote{flightlogs}, in \network{} has a Wazuh agent service that
sends events to the central database in the SOC network when a pattern within the set of
given rules is matched with an entry from one of the log sources.
The parser rules used by the Wazuh agents were selected based on a combination of
reviews, recommendations and standards, including the log policy of the Swedish Armed
Forces and SwiftOnSecurity.
586 rules from the Sigma\footnote{\href{https://sigmahq.io/}{www.sigmahq.io}}
Git repository were also included,
as well as Wazuh’s default rule set.
Sigma rules related to services not part of the scenario were excluded.
The Wazuh rules cover log data from Snort
for IP packets, Auditd
and Syslog
for Linux hosts
as well as Sysmon
and event logs for Windows systems
Every host in \network{} runs Osquery
to collect host information, such as user accounts and network interfaces.
Information about these elements are sent to Wazuh at regular intervals.

We use the Wazuh feature \enquote{Active Response} to allow defense agents to execute a
set of prepared commands on hosts in the network by calling the Wazuh REST API\@.
Two commands were implemented: one that powers off a given machine, and one that blocks
traffic between a given host and other subnets in both directions.
Traffic blocking is executed by adding new firewall rule entries to the host
\enquote{fw1}\footnote{This does not include interactions with the SOC network, to allow
the Wazuh services to report alerts and receive commands.}.
The commands are parametrized with a single argument; the identifier of the Wazuh agent
service that should execute the command. This means that a defender agent has to make two
decisions at each time step; what action to take and on what factor in the state it
should be applied on.

\subsection{Network Activity}\label{sec:crateagents}
We set up automated user agents to emulate regular work in \network{}, along with
processes such as a mail server on the host \enquote{mail}.
The simulated users follow predefined schedules in which they exchange emails, access
internal web interfaces, use remote desktop applications, and open files of various types.
Occasionally, a simulated system administrator connects to machines and executes commands
using Remote Desktop, PsExec, or \ac{wmi}.

Threat actor activity was produced using the red-team automation tool \lore{}~\cite{9740000}.
Lore attempts to compromise systems by selecting various actions, which may or may not
succeed, from a pool of available options defined by its configuration.
The initial entry point of Lore was set to the host \enquote{flightlogs}.
This host does not run a Wazuh agent, making it functionally invisible to and untouchable
by defender agents.
Lore can therefore never be fully expelled from the network, which is a similar premise
as in the Cage simulators~\cite{https://doi.org/10.1002/aaai.70021}.
Two configurations for \lore{} were used, which we call \enquote{Guided} and
\enquote{Exploratory}. These determine how \lore{} prioritizes actions and objects it discovers.
With the \enquote{Guided} configuration, \lore{} is configured with blacklists to ignore
machines that are not along the fastest path between the entry point and the
\enquote{payroll} machine in the DMZ segment.
With the \enquote{Exploratory} configuration, Lore selects machines more randomly,
and has access to more actions.
To analyze the attack paths \lore{} selects, we ran it alone in \network{} for a number
of episodes. Transition diagrams showing estimated probabilities of Lore compromising
machines, with the different configurations are shown in~\autoref{fig:loremodes} in
Appendix~\ref{sec:additionalfigure}.
\lore{} records its activity in a log stored in the Crate administration layer.
This provides a ground-truth of what actions \lore{} has succeeded and failed at during
an episode, which we use for post-hoc evaluation of the defender agents.

\subsection{Data Modeling}

We represent the state and history of the network environment as a database following a
set schema. We defined this schema using
a reduced copy of the \ac{mal} language CoreLang~\cite{KATSIKEAS2024104057}, titled
\cratelang{}\cite{buhaiu_2026_rpcde-z6r38}.
The resulting language consisted of 9 asset types,
49 attack steps and two defense steps.
The language covers two attack vectors that lead to unintended access to a host's secure
data: one through using a software vulnerability, and one where access is gained through
brute-forcing credentials.
A rendering of the attack step relations in \cratelang{} is shown in
Figure~\ref{fig:languagegraph} among the additional figures in Appendix~\ref{sec:additionalfigure}.
The language contains two defense steps, \enquote{Application\-.notPresent} and
\enquote{ConnectionRule\-.restricted},
to correspond with the two commands we implemented in the network.
Functionally, \enquote{notPresent} blocks both attack vectors for a host, making it
impossible to access its data for an attacker.
The \enquote{restricted} step blocks the associated \enquote{ConnectionRule} attack steps
representing access to the host from a different network, but still allows traffic to the
host through internal subnet connections.

\begin{figure*}[htpb]
  \centering
  \includegraphics[width=\linewidth]{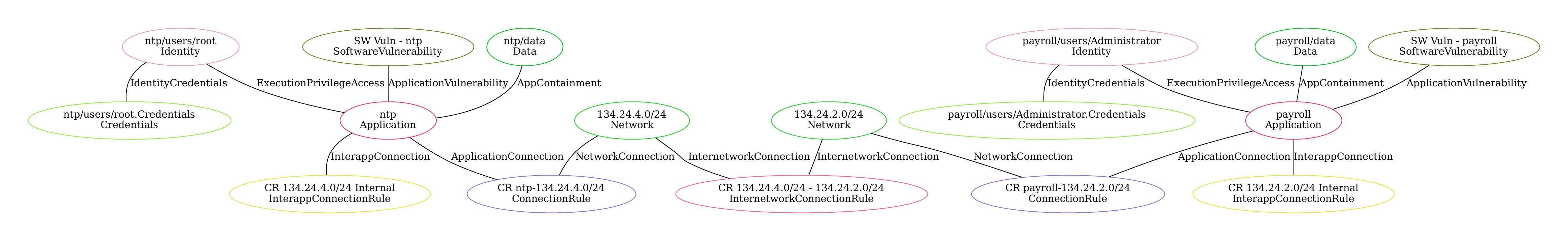}
  \caption{\ac{mal} instance model based on Osquery data from the hosts \enquote{ntp} and
    \enquote{payroll}, with asset types and relations defined in
  \cratelang{}.}\label{fig:instance-model}
\end{figure*}

We defined a procedure to construct \ac{mal} instance models from data gathered from
\hyperref[sec:network]{\network{}} with Osquery, both to create models for the simulation
and to construct observations when defender agents interface with \network{}.
The procedure fetches data from different Osquery tables
relating to hosts, users and network interfaces,
relying on shared keys to create associations between assets.
The model construction procedure is described in more detail in Appendix~\ref{sec:modelparsing}.
A graphical representation of an instance model, expressed in \cratelang{}, limited to
two hosts from \network{} is shown in \autoref{fig:instance-model}.
To connect events from Wazuh to attack steps in \cratelang{}, we defined a set of pattern
rules that maps network events to instances of attack steps.
11 attack steps were associated with separate
sets of Wazuh rule identifiers, Wazuh rule groups and rules that should be ignored.
Since attack steps are associated with assets, a requirement for a rule being included is
that it needs to contain an identifier of an asset,
such as an IP address, username or host identifier.
The alert mapping procedure is described in further detail in Appendix~\ref{sec:attackstepmappings}.
We tested the accuracy of the alert mappings by collecting data from a number of episodes
without any attacker or defender agents, but with and without user agents.
Under the assumption that all attack steps observed during these episodes were false alerts,
this yielded an average false alert probability per time step of 1.7\% per attack step
without users and 3.3\% with users.

\subsection{Cost Function}

We specify the host's security priorities in terms of confidentiality, integrity and
availability (CIA)
on a scale from 1 to 5. The specific values were provided to us by an external party also
responsible for designing the network functions.
The ratings for each host class are listed in Appendix~\ref{sec:appendixtables} in
Table~\ref{tab:cia}, and the classes of the hosts in the network is listed in
Table~\ref{tab:cia_classes}.
We use the confidentiality and integrity value for the host as a cost assigned to the defender
when the \enquote{Data.read} and \enquote{Data.write} steps for the associated asset are
registered as performed.
The defense step \enquote{Application.notPresent} incurs the availability value of the
host as a cost, and
we set \enquote{ConnectionRule.restricted} to incur half of this cost.
Our operational assumption is thus that restricting external network access to a host is
less damaging to its availability
than fully shutting it down.
The combined cost for a single time step is thus defined as
\begin{align}
  \label{eq:reward}
  r(t) = \sum_x^{\mathcal{A}(t)} [C({x}) + I(x)] + \sum_x^{\mathcal{D}(t)}A(x)
\end{align}

With \(\mathcal{D}(t)\) representing the set of active defense steps at time \(t\),
\(\mathcal{A}(t)\) the set of performed attack steps at time \(t\) and \(C\), \(I\) and
\(A\) representing mappings from an attack or defense step to a real value.
We use the \emph{return}, the discounted sum of costs, of an episode as our main
evaluation metric, calculated as
\begin{align}
  \label{eq:return}
  R = \sum_t^T r(t).
\end{align}

When interacting with \network{}, we define \enquote{Data.write} and \enquote{Data.read}
to have been performed on an asset from the time \lore{} succeeds at running commands
with elevated shell access on the corresponding host, as \lore{} was not set up with any
particular actions for emulating sensitive data being accessed.
We define a defense step as activated from the time the corresponding active response was
sent to the Wazuh REST API by a defender agent.
When using the \ac{mal} simulator, \(\mathcal{A}(t)\) and \(\mathcal{D}(t)\) are part of
the simulation, thus making them trivial to obtain.

\section{Agent Evaluation in ADS-24}\label{sec:crateeval}
The main goal of the defender agent evaluation was to measure the performance of the
defensive agents when \lore{} attacks \network{} for a set period of time.
We also test how variations to the problem, in the form of user traffic, attacker
strategies and network topology, affect the returns of the agents.
The defender agents, along with a software scaffold to fetch and send data to the cyber
range, was run on a computer with an 11th Gen Intel i7 CPU at \(4.800\) GHz and integrated graphics.

The agents based on \ac{rl} were implemented using the Vejde library and trained with the
\ac{mal} simulator. Training was done using 2 million transitions, sampled from a
combination of simulation environments using different attacker policies, attacker
entry-points and network topology variations for each episode.
The simulator and the defender agent training procedure are described in further detail
in our previous work focused on the \ac{mal} simulator~\cite{nyberg2026malsimulatorcyberoperations}.

As a comparison to the automated agents, we performed a small test where we let
five people with prior experience in log analysis perform the
same intrusion response task. Three participants
faced the \enquote{Guided} Lore policy, and two
the \enquote{Exploratory}.
All had simulated user agents present
in the network.
The participants were limited to Wazuh's Kibana dashboard for information
and a simple web form to send the predefined actions to hosts in
the network. The aggregated scores obtained by the human operators are
listed in~\autoref{tab:scores}.

\subsection{Experiment Procedure}

\begin{figure}
  \centering
  \includegraphics[width=\linewidth]{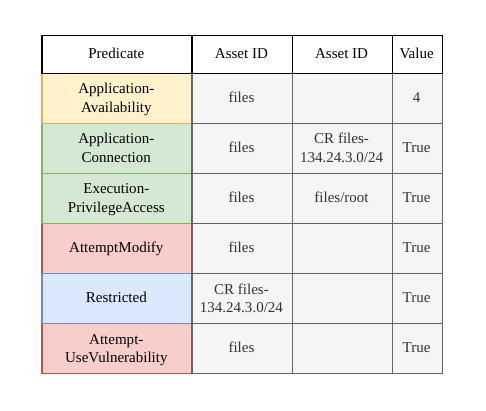}
  \caption{Example of an observation database for the defender agent, rendered as a
    table, containing facts about the host \enquote{files}.
    Predicates for attack steps are colored red, defense steps blue, asset relations green
  and CIA values yellow.}\label{fig:obstable}
\end{figure}

We evaluated the following agents: \enquote{Vejde}, reinforcement learning policy;
\enquote{Vejde w/ Noise}, same as \enquote{Vejde}, but trained with a 1\% false positive
and false negative rate per attack step; \enquote{Heuristic}, a policy that selects an
associated defense step of an asset if an associated attack step is observed;
\enquote{NoOp}, a policy that does nothing.
Each two-hour episode used a single defender agent, sampled without replacement from the
set of available agents.
The \hyperref[sec:crate]{\crate{}} snapshot functionality was used to start each episode
from the same system state.
To allow the system to settle after being restored, episodes were started half an hour
after the network was restored.
Through the course of each episode, the system maintains an \emph{observation database},
that stores past events and serve as the input to all defender agents.
The initial database is formed by network asset data collected with Osquery during the
network setup phase, and mapped to the \ac{mal} data model. The mapping procedure is
described in Appendix~\ref{sec:modelparsing}.
An example observation database rendered as a table is shown in \autoref{fig:obstable}.
An actual observation database from one of the episodes, rendered as a graph, can be seen
in Figure~\ref{fig:crate_obs} in Appendix~\ref{sec:additionalfigure}.

During the episodes, Wazuh was queried for new events every 30 seconds, and if any
returned events were matched with \iac{mal} attack step, according to the procedure
described in Appendix~\ref{sec:attackstepmappings},
an instance of the step was inserted into the observation database.
To select an action for the time step, the current database was fed to the defender
agent, producing a single action in accordance with its policy.
For the Vejde agents that produce a distribution over actions, we selected the action
assigned the highest probability by the policy. 
When there was no recorded change to the database between two time steps, no action was
requested from the current agent\footnote{Since all evaluated agents are stateless and
deterministic, the same input will yield the same action.}.
If the agent selected an action other than waiting, the action was
mapped to a corresponding active response, sent to the Wazuh server through the REST
API\@ and recorded in the database.
Assets disabled through agent actions were removed from the database along with any
associations to other assets or attack steps.
If an asset that had been removed appeared in an alert at a later time\footnote{This
  happened on occasion because of delays in the event reporting, responses being lost in
traffic and various other reasons.},
it was temporarily added to the database for a single step.
Automation of the experiments and interaction between the defender agent and \network{}
was managed by a software interface developed for the task~\cite{nyberg_2026_enqp2-38375}.

\subsection{Scenario Variations}

Experiments were run with a set of variable factors, which were selected at random before
the start of each episode.

\paragraph{Simulated Users}
To study the impact of regular network activity on the agent's performance, we ran
episodes with and without the simulated user agents, described in Section~\ref{sec:crateagents}.
Each host in the client section of \network{} is assigned a simulated user agent, which
will perform actions based on a given policy.

\paragraph{Attacker Strategy}
To test how the defender agents handle different attack profiles, we used the two \lore{}
configurations, as described in Section~\ref{sec:crateagents}.
The configuration determines which machines Lore prioritizes.

\paragraph{Network Topology}
To test how the agents handle variations to the network topology, we
randomly remove hosts, selected from \enquote{rootca}, \enquote{timereporter} and
\enquote{print}, from \network{} before the episode.
These machines are not part of the list of machines Lore is directed at with the
\enquote{Guided} policy.

\section{Results}

\begin{figure*}[tpbh]
  \centering
  \subcaptionbox[c]{Returns with no user agents.\label{fig:violinwithnousers}}
		{\includegraphics[width=\linewidth]{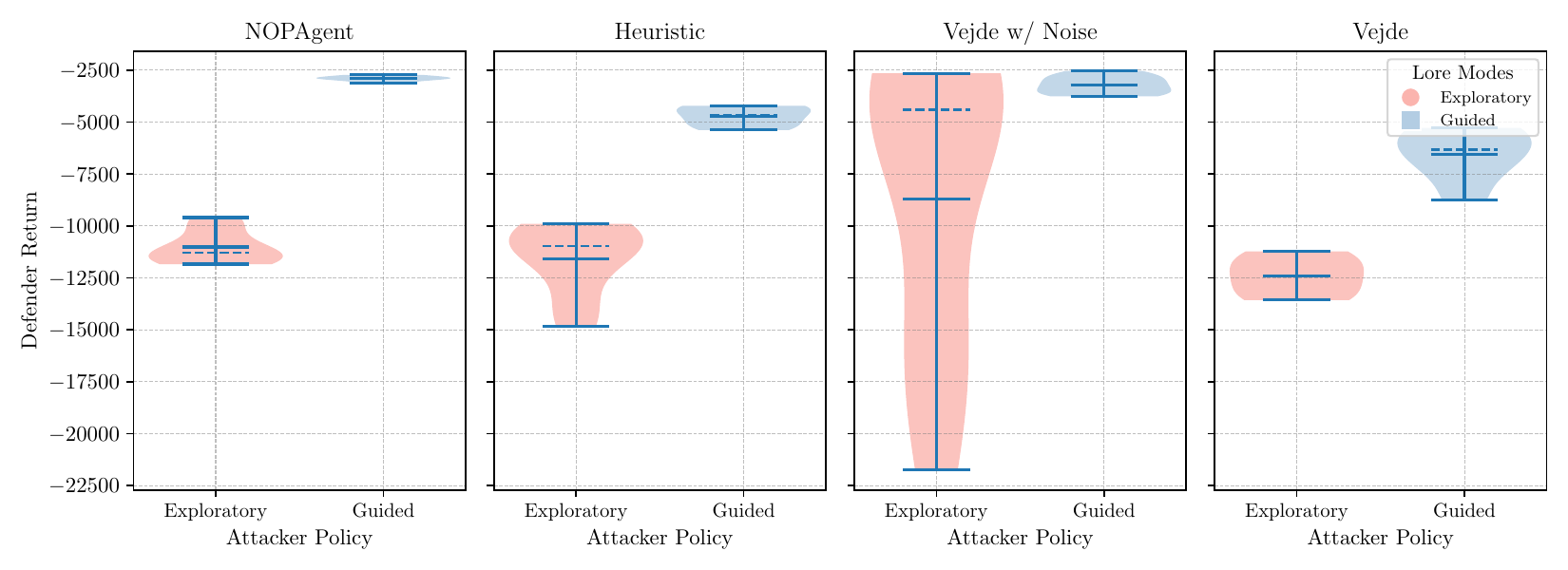}}
  \subcaptionbox[c]{Returns with user agents.\label{fig:violinwithusers}}
		{\includegraphics[width=\linewidth]{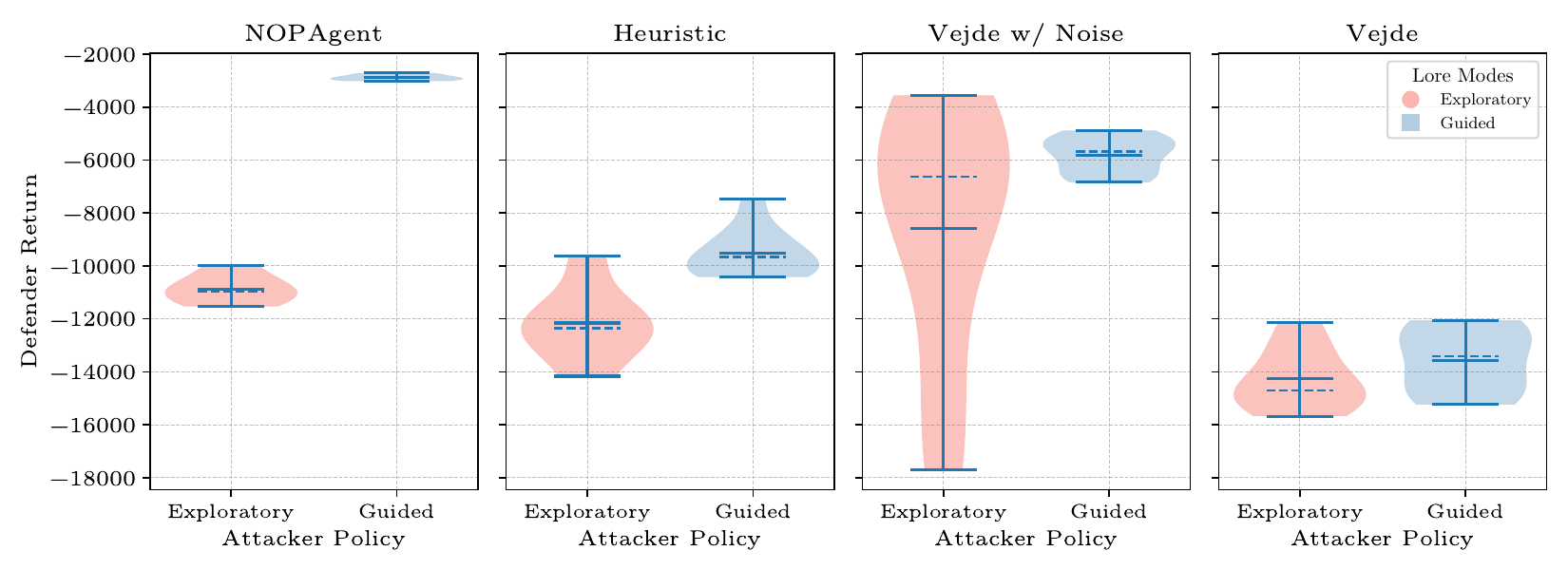}}
  \caption{Violin plots of return distributions from running defender agents in \network{} against \lore{}, with and without user agents present.
		  Higher returns indicate desired behavior.
		Median value indicated as dashed line.}\label{fig:crateresults}
\end{figure*}

Our results are based on data gathered from 134 episodes of the attack-defense scenario, with \(\approx 30\) episodes per agent type. 
All logs that we gathered can be accessed via the KTH Data Repository~\cite{nyberg_2026_4mh6z-zw065}.
Among the evaluated defender agents, the \enquote{Vejde w/ Noise} agent received the highest overall average return.
This largely comes from the agent receiving lower defense costs than the \enquote{Vejde} and \enquote{Heuristic} agents.
The lower defense cost comes from not disabling as many assets, thus inducing a lower availability cost. 
The more lenient strategy of the \enquote{Vejde w/ Noise} led to it letting \lore{} access the client net in some episodes, thus inducing higher costs.
The \enquote{Vejde} agent has the lowest overall attack cost, but also the highest average defense cost. 
The human operators were the best at avoiding false alerts, thus keeping the overall cost low.
This likely comes from the \enquote{Vejde} agent acting similarly to the \enquote{Heuristic} agent, but with the additional capability of being able to act based on events from associated assets, whereas the \enquote{Heuristic} policy will only act on assets that are directly affected by attack steps.
This more aggressive strategy is reflected in the distribution of agent actions, where the \enquote{Vejde} policy has the lowest overall probability of waiting, at 85\%, compared to the \enquote{Heuristic} policy at 88\%.
The \enquote{Vejde w/ Noise} has an overall 94\% probability of waiting, indicating that it is more conservative than the two other agents.
Violin plots of the defender agent returns can be seen in \autoref{fig:crateresults}, with and without simulated users respectively.
A table of all agent scores, also separated into attack and defense cost, can be seen in~\autoref{tab:scores}.
A scatter plot of the same returns over time can be seen in Figure~\ref{fig:scoreovertime} in Appendix~\ref{sec:additionalfigure}.

\subsection{Impact of Experimental Factors}

To analyze the impact of the different experimental factors on the returns, we ran an ANOVA analysis on the returns of each agent.
The choice of Lore's policy, being either \enquote{Guided} or \enquote{Exploratory} was the most significant variable affecting the returns of all agents.
The removed machines did not significantly affect the returns of any agent.
The user agents were significant for most defender-attacker-policy combinations, though not for the combinations (\enquote{Heuristic}, \enquote{Exploratory}) and (\enquote{Vejde w/ Noise}, \enquote{Exploratory}).
This lack of change may be caused by Lore scanning and attempting connections to machines in the client net when using the \enquote{Exploratory} strategy, even if it can not capture any machines.
This generates alerts in the client net, similar to when the simulated users are present.
As Lore's initial access machine, \enquote{flightlogs}, is not under the control of the defender agent, the defender can not do much to prevent this apart from blocking connections in the client net, incurring costs as a consequence.
Though the changes were not always significant, all decision-taking agents receive lower average returns when users are present.
This is primarily caused by users generating false alerts that cause the agent to act.
Against \enquote{Guided}, returns with users drop by \(-80\%\) for \enquote{Vejde w/ Noise}, \(-101\%\) for \enquote{Heuristic} and \(-109\%\) for \enquote{Vejde}.
The drop is mainly caused by increased defense costs, but there is also an increase in the attack cost for \enquote{Vejde} and \enquote{Heuristic} with users.
This may be caused by \lore{} getting more time to act while the defender agent handles false alerts.
As mentioned, users make less of an impact when facing \enquote{Exploratory}, with a drop of \(-5\%\) for \enquote{Heuristic}, \(-11\%\) for \enquote{Vejde w/ Noise} and \(-14\%\) for \enquote{Vejde}.

All agents receive higher average returns against the \enquote{Guided} configuration than against the \enquote{Exploratory},
owing to \lore{} not exploring as many machines in the former.
When faced with this strategy, the do-nothing \enquote{NoOp} strategy actually receives the highest average return of all agents, both with and without users.
Though somewhat counter-intuitive, as the return is entirely composed of the attack cost from \lore{} compromising machines,
the scenario we use prioritizes availability, meaning that defenses are expensive to use unless needed.
Under the circumstances that \enquote{Guided} creates, where \lore{} does not compromise many valuable machines, a defender agent thus needs to be precise with its actions.

\subsection{Alternate Cost Balances}

Given that the return is calculated as a sum of attack and defense costs, \(R = R_a + R_d\), we calculated different \(R_\alpha = (1-\alpha)R_a + \alpha R_d\) for with \(\alpha \in [0, 1]\).
At \(\alpha=0\), we only care about the availability of the system.
At this point, the \enquote{NoOp} agent is always the best choice among the agents, as all of our defender actions impact availability in some way.
On the other end where availability does not matter, at \(\alpha=1\), the \enquote{Vejde} policy is the best choice for most configurations.
The intersection between a line formed by an agent's returns and the line formed by the \enquote{NoOp} agent represent the point at which it would be better to use the agent compared to NoOp.
From the diagrams, it can be seen that against the \enquote{Guided} policy, the attack cost would need to be roughly 30\% higher for any of the agents to be viable, whereas against the \enquote{Exploratory} policy the opposite is true.
Since \enquote{Exploratory} induces a higher attack cost for waiting, we could lower the attack cost by 30\%, and \iac{rl} agent would still be better than the \enquote{NoOp} agent.
The diagrams showing the scores for different values of \(\alpha\) can be seen in Figure~\ref{fig:balance} among the additional figures of Appendix~\ref{sec:additionalfigure}.

\begin{table*}[htpb]
  \centering
  \caption{Mean returns, attack costs and defense costs with standard deviation for defender agents when evaluated in \network{}.
		Note that the return is the sum of the attack and defense costs.}\label{tab:scores}
		\begin{tabular}{llcl|ll}
		  \toprule
		  Agent Type & Attacker Policy & Users Present & Return & Attack Cost & Defense Cost \\
		  \midrule
		  NoOp & Exploratory & False & -11020 ± 826 & -11020 ± 826 & 0 ± 0 \\
		  Heuristic & Exploratory & False & -11586 ± 1737 & -1344 ± 360 & -10242 ± 1520 \\
		  Vejde & Exploratory & False & -12422 ± 955 & -918 ± 227 & -11503 ± 1037 \\
		  Vejde w/ Noise & Exploratory & False & -7779 ± 5723 & -3816 ± 3218 & -3963 ± 2615 \\
		  \midrule
		  NoOp & Guided & False & -2904 ± 133 & -2904 ± 133 & 0 ± 0 \\
		  Heuristic & Guided & False & -4721 ± 514 & -931 ± 125 & -3790 ± 417 \\
		  Vejde & Guided & False & -6307 ± 1332 & -816 ± 377 & -5491 ± 1149 \\
		  Vejde w/ Noise & Guided & False & -3214 ± 429 & -1673 ± 387 & -1541 ± 311 \\
		  \midrule
		  \midrule
		  NoOp & Exploratory & True & -10875 ± 546 & -10875 ± 546 & 0 ± 0 \\
		  Heuristic & Exploratory & True & -12157 ± 1467 & -2050 ± 662 & -10107 ± 1040 \\
		  Vejde & Exploratory & True & -14270 ± 1181 & -1815 ± 454 & -12456 ± 1362 \\
		  Vejde w/ Noise & Exploratory & True & -8597 ± 4745 & -3032 ± 2150 & -5565 ± 2789 \\
          
		  \midrule
		  NoOp & Guided & True & -2883 ± 117 & -2883 ± 117 & 0 ± 0 \\
		  Heuristic & Guided & True & -9525 ± 929 & -1479 ± 477 & -8046 ± 946 \\
		  Vejde & Guided & True & -13205 ± 1569 & -1435 ± 590 & -11770 ± 1291 \\
		  Vejde w/ Noise & Guided & True & -5812 ± 694 & -1655 ± 122 & -4157 ± 735 \\
          \midrule
          \midrule
          Human & Exploratory &  True & -2875 ± 365 & -1806 ± 356 & -1069 ± 8 \\
        Human & Guided &  True & -2956 ± 564 & -2141 ± 497 & -815 ± 215 \\
		  \bottomrule
		\end{tabular}
\end{table*}

\section{Discussion}

The results from our test suggests that the \enquote{Vejde w/ Noise} \ac{rl} agent is a
better option for defending \network{} against \lore{} than the simple heuristic agent in
terms of minimizing the joint attack and defense cost.
This shows that our relational \ac{rl} approach functions in the realistic environment,
and supports our hypothesis that \iac{rl} agent can outperform the manually-crafted
heuristic policy at the defense task we have set up.
This is a promising result for the application of \ac{rl} agents for use in cyber attack
mitigation, especially given how simplified our simulation and data model is in relation
to the real system.
However, we emphasize that the current approach is still
far from being practical to a real organization,
namely in how sensitive the approach is to false alerts.
The problems that need to be addressed are not only technical,
but also operational.
The introduction of automated agents of any type may for instance create new threat
surfaces for adversaries to exploit.
Additionally, the defense task has been heavily simplified and encoding complex
operational needs into a scalar cost function may fail to capture higher-level demands
from users and operators.

\subsection{Agent Explainability \& Interpretability}

While we can analyze the actions of the \ac{rl} agents, we can not directly interpret
what behaviors their policy functions have encoded.
Explainable and interpretable agents and policies are important if they are to be used
for automation in security-critical contexts~\cite{10.1145/3729213}.
In the context of intrusion detection, some authors have suggested to use simpler model
architectures, such as decision trees, to check datasets and trained
models~\cite{10.1145/3548606.3560609, 10629000} for unintended biases.
This can reveal if the agent is overly reliant on a feature that would be unrealistic in
a different context.
Decision trees can be trained to mimic the output of the neural network classifier model,
and a domain expert can then analyze to tree to determine whether the biases and features
used by the model are justifiable given the problem.
We did attempt to train a decision tree to mimic the output of a Vejde agent.
In our simple implementation however, the input to the decision tree was fully grounded.
As such, the resulting tree does not fully represent the inductive policy parametrized by
the \ac{gnn}.
There may be a way of extracting an interpretable policy, as either a decision list or
tree, by for instance inductive logic programming, but we considered it out of scope for this paper.

\subsection{On Replicability and Reproducibility}

We recognize that independently replicating a study like this is difficult, since access
to \network{} and \lore{} can not be shared publicly.
For the sake of transparency, we publicly share all data that was included in our
evaluation, which includes events recorded by Wazuh, events parsed by the monitor, the
activity log of \lore{} and the activity log of the simulated users.
We have made the source code for several of the components public.
This includes the \ac{mal} simulator and the scenario files used for training the agents,
as well as the interface for mapping Wazuh data to \iac{mal} model.
The interface can be run offline with limited functionality\footnote{Actions do not cause
effects when running on saved data, for instance.} using saved Wazuh data.

\subsection{Training from network data}

We use a model-free \ac{rl} method to train our agents, which means that we can
theoretically train them directly against \network{}.
Previous work has demonstrated this is possible, albeit with
much smaller networks and a significant time investment~\cite{ITURBE2026129168}.
We use a simulator to train agents before deploying them in the evaluation network.
This introduces risks of performance deterioration due to differences
between the dynamics of the simulator and the target system.
A possible fusion of these approaches is to perform pre-training in the simulator,
and then fine tune the policy using data from the real network.
We may also use already collected data to pre-train the agent
using offline \ac{rl} methods. In evaluating both of these
possible approaches, the amount of samples required for the policy
to converge is a primary evaluation metric in addition to the
cost function.

\section{Conclusion}

We address issues present in previous work on automated network intrusion mitigation that
assume a fixed network topology, by utilizing a combination of relational data modeling
and graph representation learning. We have evaluated the approach in a cyber range
of networked virtual machines.
The scale of our evaluation, in terms of network size and time frames,
is larger than any related work we were able to find.
The defender agent trained using relational \ac{rl} with added measurement noise induced
the lowest overall average cost when defending the network against a red-team emulation tool.
The cost was highly dependent on the attacker policy,
and to a lesser extent the presence of user agents.
Though our results support the thesis that \ac{rl} may be used for automated cyber operations,
several challenges remain before practical application would be possible.
Evaluating agents in large-scale emulated network environments is work-intensive and
prone to a variety of technical issues, but we believe such evaluations are important to
test the claim that machine learning can be used for cyber security tasks.
We hope that by sharing our experiences we may facilitate such evaluations in future works.

\section{Acknowledgments}

We would like to thank Matyas Barocsai, Kristoffer Lundholm and Jonas Almroth at \ac{foi}
for setting up and maintaining the network in \crate{}.
We would also like to thank Fredrik Erling and Mathias Bjarme at \ac{fmv} for creating
the original scenario with the network topology and CIA values.

\section{CRediT author statement}
\textbf{Jakob Nyberg:} Conceptualization, Methodology, Software, Validation, Formal
analysis, Investigation, Data Curation, Writing --- Original Draft, Writing --- Review \&
Editing, Visualization
\textbf{Teodor Sommestad:} Software, Resources, Supervision, Writing --- Review \& Editing.
\textbf{Mathias Ekstedt:} Conceptualization, Supervision, Writing --- Review \& Editing,
Project administration, Funding acquisition.
\textbf{Andrei Buhaiu:} Software, Methodology.
\textbf{Joakim Loxdal:} Software, Writing --- Review \& Editing.
\textbf{Pontus Johnson:} Supervision, Writing --- Review \& Editing.

\printbibliography%

\clearpage

\appendix

\section{Additional Implementation Details}\label{sec:appendiximplement}

\subsection{MAL Instance Model from Osquery Data}\label{sec:modelparsing}

\hyperref[sec:mal]{\ac{mal} instance models} of the network were constructed from data gathered from \hyperref[sec:network]{\network{}} with Osquery, both to create models for the simulation and to construct observations during inference in \network{}.
The data consisted of the Osquery tables \enquote{users}, \enquote{system\_info}, \enquote{os\_version}, \enquote{interface\_details} and \enquote{interface\_addresses}.
Each of the Wazuh agents in \network{} produces one instance of each table, and if an agent is missing one of the tables it is assumed to be missing and disregarded from the model.
Certain asset types, such as \enquote{SoftwareVulnerability} assets, need to be present in the model for the \cratelang{} attack graph to be traversable to attackers in the \ac{mal} simulator.
However, as we did not have a practical method of mapping these assets to Osquery data, they are inferred based on the presence of other assets.
We also did not have an in-network data source to instantiate \enquote{Network} assets to represent the network segments, so these are added the model generation process through a secondary data source\footnote{Specifically, a YAML file that is loaded with the model generator.}.
We consider this an acceptable workaround since the base firewall configuration for the network is assumed to be static throughout the entire evaluation.

One \enquote{Network} asset is created for each CIDR range in \network{}.
The ranges and their pairwise connectivity are supplied from a secondary data source.
Networks were associated through \enquote{InternetworkConnectionRule} assets to mirror the connectivity in \network{}.
We also add an \enquote{InterAppConnectionRule} asset for each CIDR range.
This models that hosts can communicate inside the network section, even if the \enquote{ConnectionRule} asset for the host is restricted.
The \enquote{InterAppConnectionRule} asset is associated with every \enquote{Application} asset representing a host in the IP range.
One \enquote{Application} asset is created for each agent to represent the host operating system.
Each \enquote{Application} is associated with inferred \enquote{SoftwareVulnerability} asset and an inferred \enquote{Data} asset.
Each interface in the \enquote{interface\_addresses} table instantiates one \enquote{ConnectionRule} asset to model network connectivity.
\enquote{ConnectionRule} assets are associated with the respective \enquote{Application} representing the host the interface belongs to,
as well as the \enquote{Network} asset representing the network that the host belongs to.
\enquote{Identity} assets are parsed from the \enquote{users} table to model user accounts.
We only model one user per host, and select the administrator username as the asset identifier.
Each \enquote{Identity} is associated with the respective \enquote{Application} asset representing the host the user belongs to, and an inferred \enquote{Credentials} asset.

\subsection{Wazuh Rule to Attack Step Mapping}\label{sec:attackstepmappings}
In the \ac{mal} simulator, there is a direct mapping between the attack steps performed by the attacker and the observations provided to the defending agent.
This is not true when interacting with \hyperref[sec:network]{\network{}}, as we can only observe the state through the lens of Wazuh logs.
As such, as mapping from Wazuh rules to \ac{mal} attack steps was needed.

Wazuh contains a set of default rules, which we supplemented with a set of rules from Sigma.
Classification of Wazuh rules to \cratelang{} attack steps was done manually based on alert data collected from running \lore{} and simulated users for a week in \network{}.
We acknowledge that this introduces a degree of bias to the process,
in that we know the set of triggered alerts from the significantly larger set of enabled Wazuh rules.
However, the collected data contained not only alerts generated by \lore{},
but also the simulated users as well as services running on hosts in \network{}.
As such, the resulting mapping prioritizes recall, not precision, as most alerts that were observed were mapped to an attack step regardless of its origin.
Each attack step is associated with a set of rule identifiers, a set of rule groups and a set of rule IDS that should be ignored.
If a Wazuh rule matches either one of the rule identifiers or the rule group, while the rule ID is not in the set of ignored rules, the rule is mapped to a corresponding attack step.
For some attack step types, we add additional data-dependent matching rules.
For instance, for the attack step type \enquote{attemptConnectToApplication}, alerts with different source and destination subnets in their data fields are mapped to the \enquote{ConnectionRule} instance of step, and those with the same are mapped to the same attack step in \enquote{InterAppConnectionRule}.
The full set of rule mappings can be seen in the Git repository of the monitor~\cite{nyberg_2026_enqp2-38375}.

\clearpage

\subsection{Experiment Anecdotes}

This subsection contains a number of anecdotes of incidents we encountered while running the experiments, to illustrate the various kinds of issues one might encounter while running experiments using a cyber range.
We include these for others to hopefully learn to avoid the mistakes we have made along the way.

\paragraph{Load-bearing User Accounts}

We initially included a defender action for removing user accounts.
However, Lore primarily compromises the root user account, and removing this breaks the machine.
We could not remove users from hosts as lore primarily takes the root user and removing this breaks the machine.

\paragraph{Kill Unconfirmed}

We assume that defender actions always succeed. However, this was not always the case. Active Responses would sometimes fail to execute due to Wazuh event queues being full, for instance.
It would be more robust to query the network for confirmation. This is made difficult, however, by the fact
that some of the actions turn machine off.



\paragraph{Defanged Attacker}

For a number of experiment rounds, issues with \lore{} caused it to only.
As we were running with only the \ac{rl} agents and the heuristic agent at the time, the poor performance of \lore{} was attributed to being blocked by the defender.
We only discovered this issue by comparing our agents with the NOP agent, as doing nothing against an adversary that does nothing is a very good strategy.
When then used the first interval of each day as a calibration round, where \lore{} was run without a blue agent.
This allowed us to continually confirm that \lore{} was succeeding with its attacks when not interrupted.
This emphasizes that using simple baselines is important to gauge the difficulty of the task and debug the system.

\paragraph{See no Evil}

Defender actions can have unintended side effects.
One of the active responses is to block a machine from communicating with other subnets.
In our first version of this action, the machine was blocked from communicating with \emph{all} other subnets.
In practice, this meant that the block action would also block the Wazuh agents from sending alerts to the SOC subnets, effectively blinding the agent to all activity on the machine.

\paragraph{Unsafe Users}
Ironically, one of the biggest threats to the network security was not \lore{}, but the simulated user agents.
In their first configuration, the agents were equipped with the ability to turn on machines, as this is reasonable for a human user to do.
However, this meant that the users would effectively counter the defender agent, and turn machines back on after the defender had turned them off.
While this behavior is reminiscent of actual users with poor security training, we ultimately decided to disable this feature of the user agents.

\paragraph{Back to Zero}
At one point, the entire network topology was wiped due to a technical mishap. Luckily, snapshots were available but were missing some later additions, which had to be restored manually. 
This highlights the need to keep backups of the system, or up-to-date declarative definitions of the system so that it can be easily restored.



\paragraph{Background Radiation}

Even without any agents in the net, we observed events being generated in \network{}.
These were usually caused by Windows processes such as Windows Defender, or Microsoft Exchange running on the \texttt{mail} host.

\paragraph{Experiment Automation}

As running the experiments was a time-consuming process,
most of the work with starting and running the experiments was done automatically by various runners scripts.
At midnight, the virtual machine running the defender agents was rebooted, and a schedule of agents to run for the day was generated.

\paragraph{Missing Wazuh Agent}

While \enquote{flightlogs} was always intended to be an entrypoint machine for Lore,
we intended it to run a Wazuh agent just like any other machine in \network{}.
Due to a technical mistake that went unnoticed for the duration of the evaluation, flightlogs was run without a Wazuh client.
This means that Lore's presence can not be fully removed from the network, and the events that occur in flightlogs were invisible to the defender agents.

\paragraph{Accidental Persistence}
In the \ac{mal} simulator, the attacker can be fully blocked by defenses, removing all possible actions for it.
This threat model can be contrasted with the Cage simulations~\cite{https://doi.org/10.1002/aaai.70021} where the attacker can never be fully expelled, and will continue taking actions during the entire episode.
We had intended the scenario to work like the former model in \network{}, to match the MAL Simulator, but due to the missing Wazuh agent we wound up with a scenario more similar to the latter.
This meant that even if Lore was blocked from accessing other machines, it could still run certain actions, like ping scans, from \enquote{flightlogs}.

\paragraph{Missing Attack Step}
We did not attempt to estimate false negative rates, 
as this would have required an additional mapping from Lore's actions to \ac{mal} attack steps.
Due to a technical mishap however, the attack step \enquote{ConnectionRule.attemptAccessNetworks}, 
indicating connections to other subnets, was not matched with any events during the course of the evaluation, functionally setting its false negative rate to 1.0 in \network{}.

\clearpage

\subsection{Additional Tables}\label{sec:appendixtables}
\begin{table}[!htpb]
  \centering
  \caption{Hyperparameters used for training Vejde agents.}\label{tab:hyperparams}
  \begin{tabular}{c|c}
    \toprule
    RL & Value \\
    \midrule
    Maximum Episode Length & 300 \\
    Minimum Episode Length & 120 \\
    Epochs per Batch & 8 \\
    Number of Parallel Environments & 16 \\
    Discount Factor & 0.99 \\
    GAE \(\lambda\) & 0.95 \\
    Value Function Loss Coefficient & 0.1 \\
    Weight Decay & \(10^{-2}\) \\
    \midrule
    Vejde & \\
    \midrule
    Message Passing Steps & 4 \\
    Activation Function & \(\tanh\) \\
    Aggregation Function & \(\sum\) \\
    \midrule
    Warmup & \\
    \midrule
    Learning Rate & \(10^{-3}\) \\
    Entropy Coefficient & \(10^{-2}\) \\
    Maximum Gradient Norm & \(1.0\) \\
    Batch size/Rollout length & \(1024\) \\
    Minibatch size & 1024 \\
    PPO Clipping Fraction & 0.2 \\
    \midrule
    Finetune & \\

    \midrule

    Learning Rate & \(10^{-4}\) \\
    Entropy Coefficient & \(10^{-4}\) \\
    Maximum Gradient Norm & \(0.1\) \\
    Batch size/Rollout length & \(2048\) \\
    Minibatch size & 1024 \\
    PPO Clipping Fraction & 0.1 \\
    \bottomrule
  \end{tabular}
\end{table}
\newpage
\begin{table}[!htpb]
\caption{Confidentiality, integrity and availability priorities for different assets in the network.}
\begin{tabular}{lccc}
\toprule
Host Class & Availability & Integrity & Confidentiality   \\
 \midrule
Time Server / NTP&  4 &  5&  1 \\
Log Server&  5&  5&  4 \\
File Server&  4&  2&  2 \\
Domain Controller&  5&  5&  2\\
Name Server&  5&  5&  2\\
Web Server&  2&  4&  1 \\
CA-Server&  5&  5&  1 \\
Clients&  5&  5&  2\\
Mail Server&  3&  4&  3\\
Mail Relay&  3&  4&  3\\
Payroll Server&  5&  5&  2\\
DB Server&  5&  5&  2\\
\bottomrule
\end{tabular}\label{tab:cia}
\end{table}
\begin{table}[!htpb]
\caption{Classes of host that can appear in the network.}
\centering
\begin{tabular}{lc}
\toprule
Hostname & Type   \\
\midrule
timereporter & Time Server / NTP   \\
ntp & Time Server / NTP \\
files & File Server      \\
dc & Domain Controller                 \\
weborder & Web Server              \\
mail & Mail Server \\
mailrelay & Mail Relay \\
payroll & Payroll Server \\
print & Mail Server \\
ns & Name Server \\
db & DB Server \\
ca & CA-Server                  \\
flightlogs & Log Server \\
ap[1--16] & Clients \\
\bottomrule
\end{tabular}\label{tab:cia_classes}
\end{table}



\begin{table}[htpb]
    \centering
    \caption{Human Operator Scores}
\begin{tabular}{lrrrl}
\toprule
Participant & Defense Cost & Attack Cost & Total Cost & Lore Policy \\
\midrule
A & -804 & -1568 & -2372 & Guided \\
B & -1035 & -2462 & -3497 & Guided \\
C & -606 & -2393 & -2999 & Guided \\
D & -1075 & -2058 & -3133 & Exploratory \\
E & -1063 & -1554 & -2617 & Exploratory \\
\bottomrule
\end{tabular}
\end{table}

\onecolumn
\subsection{Additional Figures}\label{sec:additionalfigure}
\begin{figure*}[tbph]
\centering
\subcaptionbox[c]{Returns with no user agents.\label{fig:scoreovertimenousers}}
{\includegraphics[width=\linewidth]{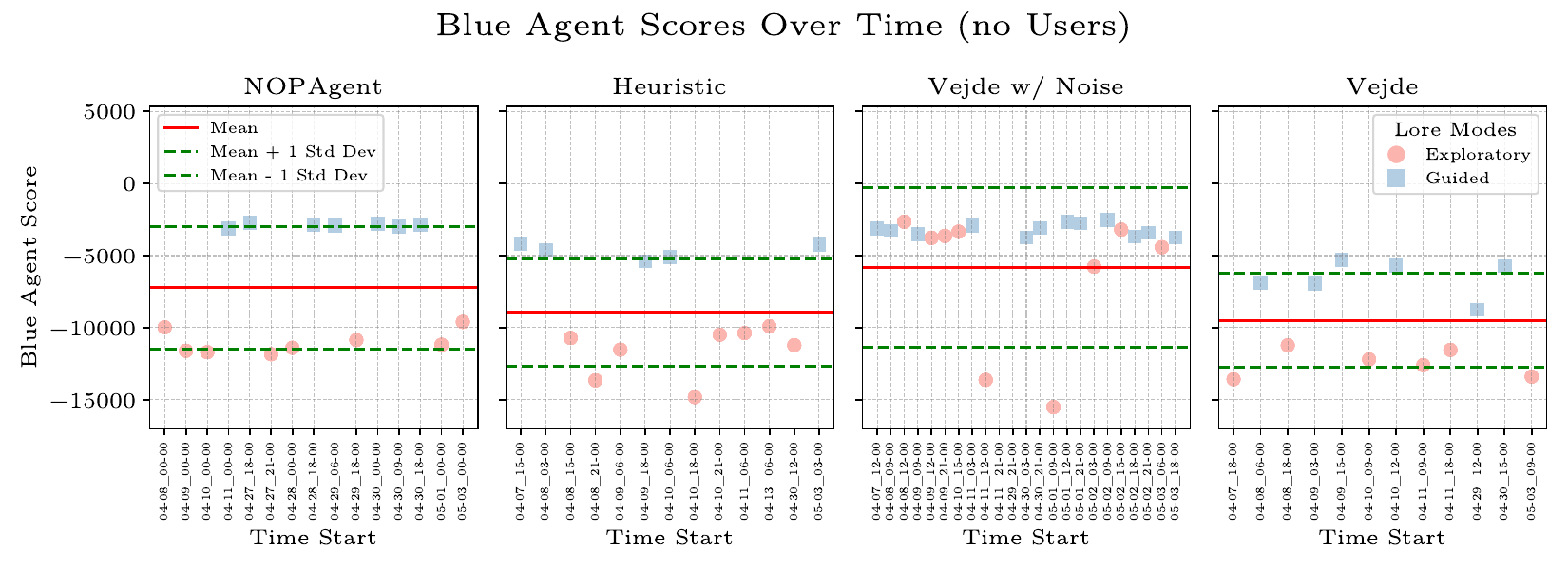}}
\subcaptionbox[c]{Returns with user agents.\label{fig:scoreovertimeusers}}
{\includegraphics[width=\linewidth]{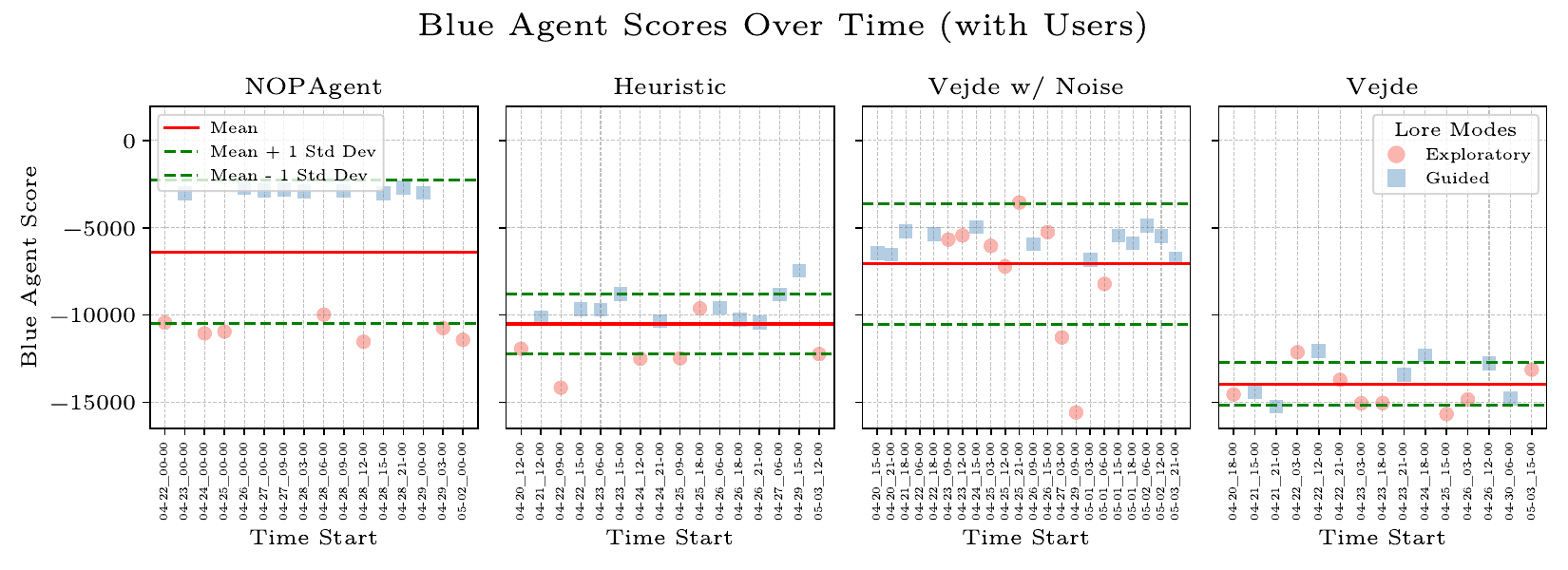}}
\caption{Scatter plots of returns over time after running defender agents in \network{} against \lore{}, with and without user agents present. Higher returns are better.}\label{fig:scoreovertime}
\end{figure*}
\begin{figure*}[!htpb]
    \includegraphics[width=\textwidth]{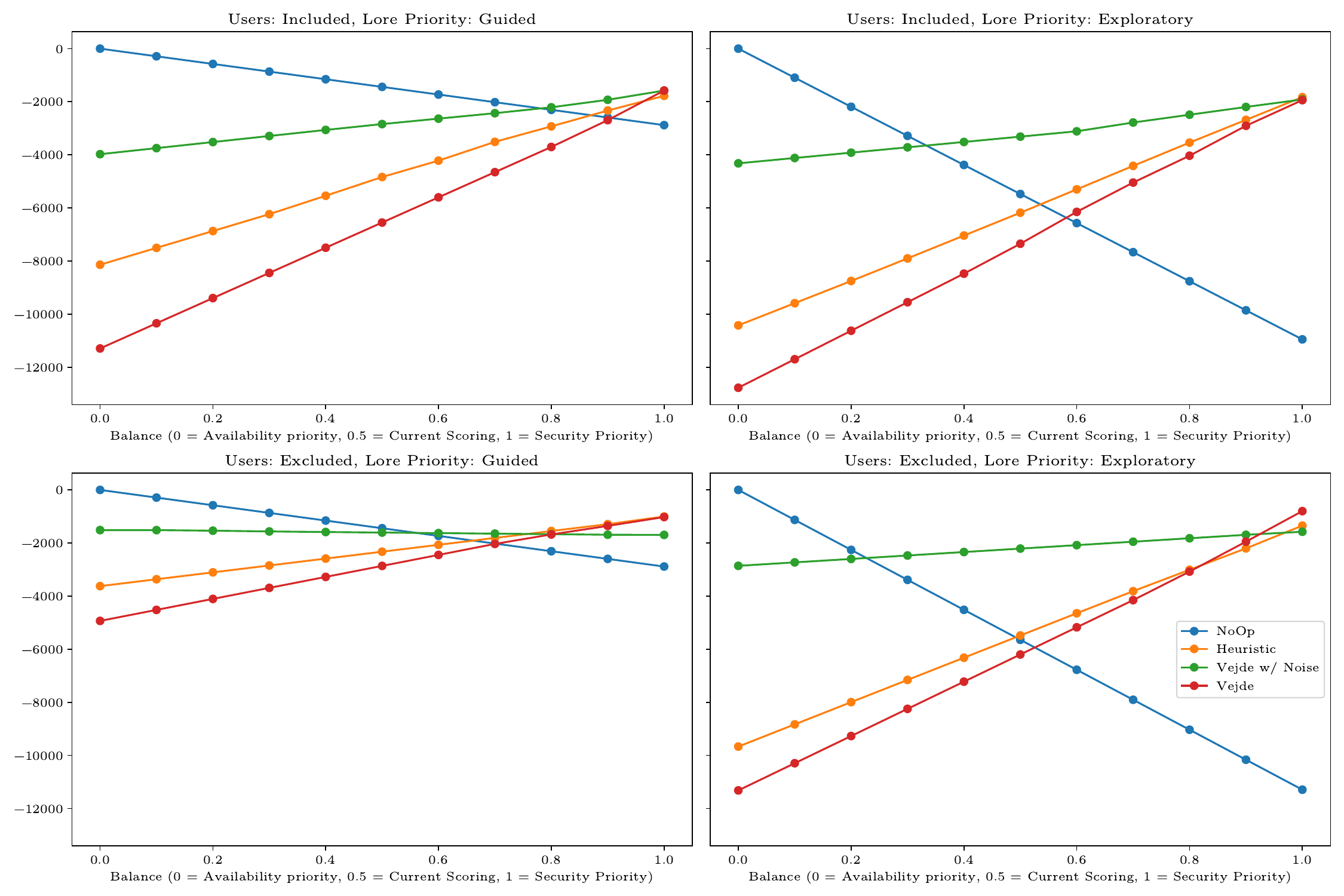}
    \caption{Agent returns for different prioritizations of attack and defense cost, calculated as \(R_\alpha = (1-\alpha)R_a + \alpha R_d\), with \(\alpha=0.5\) representing the current balance.}\label{fig:balance}
\end{figure*}



\begin{figure*}[!htpb]
    \centering
    \includegraphics[width=\textwidth]{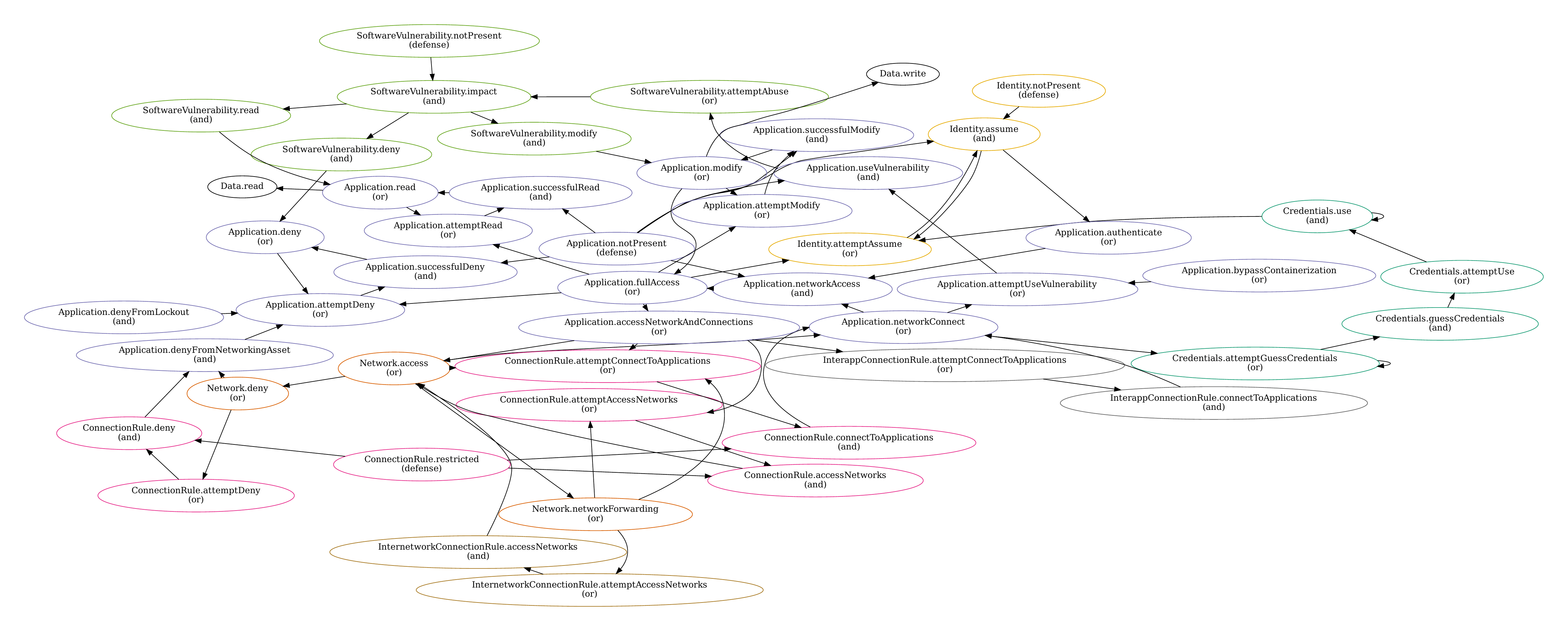}
    \caption{Graph depiction of the \ac{mal} \cratelang{} language. Nodes represent attack steps, with directed edges showing possible subsequent steps.}\label{fig:languagegraph}
\end{figure*}

\begin{figure*}[!htpb]
  \centering
\begin{subcaptionblock}{0.5\textwidth}
\centering
\includegraphics[width=0.8\linewidth]{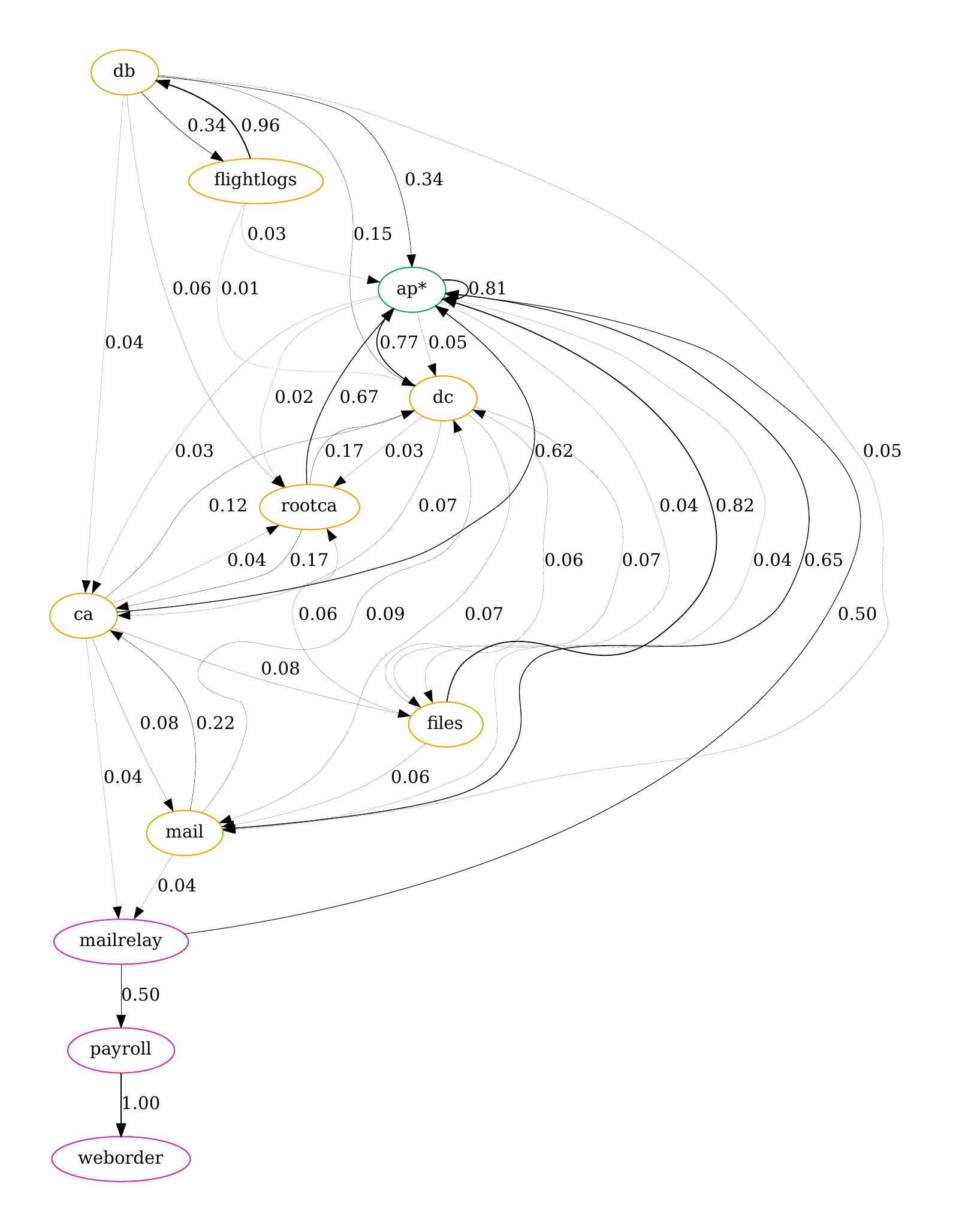}
\caption{Exploratory.
  Hosts in the client net have been combined into a single node for clarity.}
\end{subcaptionblock}%
\begin{subcaptionblock}{.5\textwidth}
\centering
\includegraphics[width=0.8\linewidth]{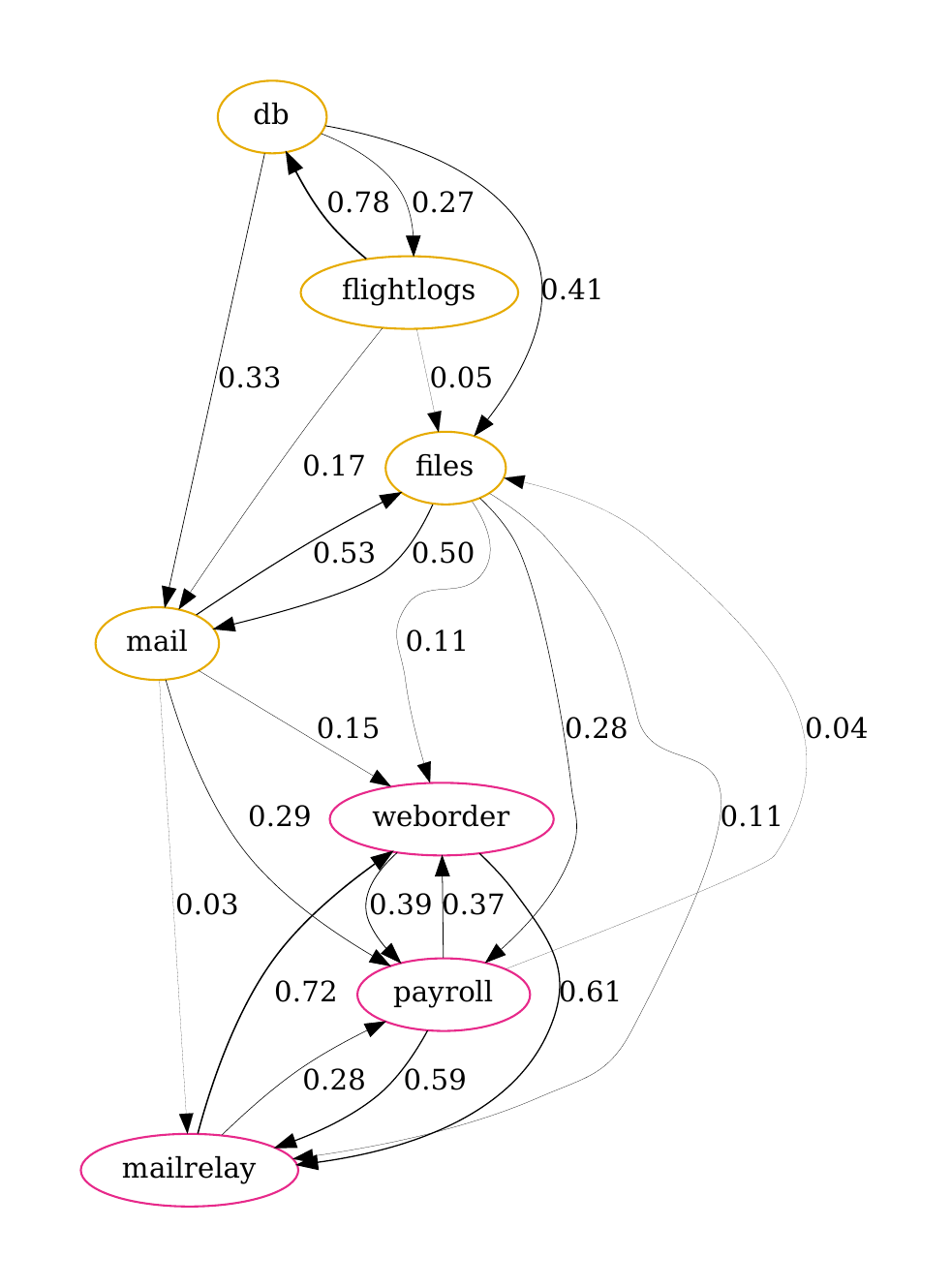}
\caption{Guided.
  With this configuration, \lore{} does not compromise any of the hosts in the client net.}
\end{subcaptionblock}%
\caption{Graphs showing estimated probability of \lore{} compromising a host when not interrupted with the two strategies, \enquote{Guided} and \enquote{Exploratory}.
  Probabilities \(<\) 1\% are not drawn.}\label{fig:loremodes}
\end{figure*}

\begin{figure*}[htpb]
    \centering
    \includegraphics[width=1.25\textwidth, angle=90]{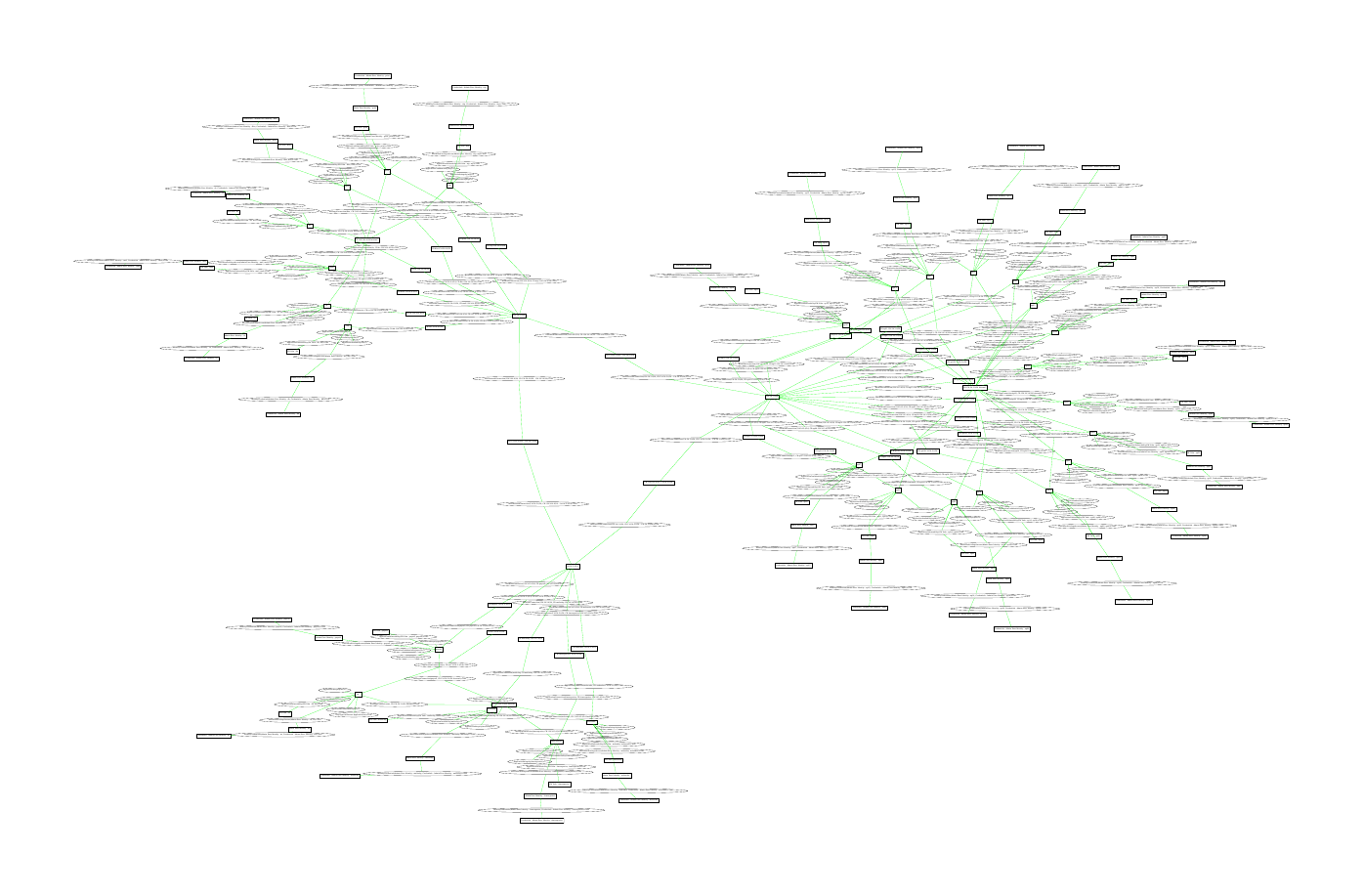}
    \caption{Bipartite graph representation of an initial MAL Simulator state, converted to the Vejde representation format. }\label{fig:sim_obs}
\end{figure*}
\begin{figure*}[htpb]
    \centering
    \includegraphics[width=1.25\textwidth, angle=90]{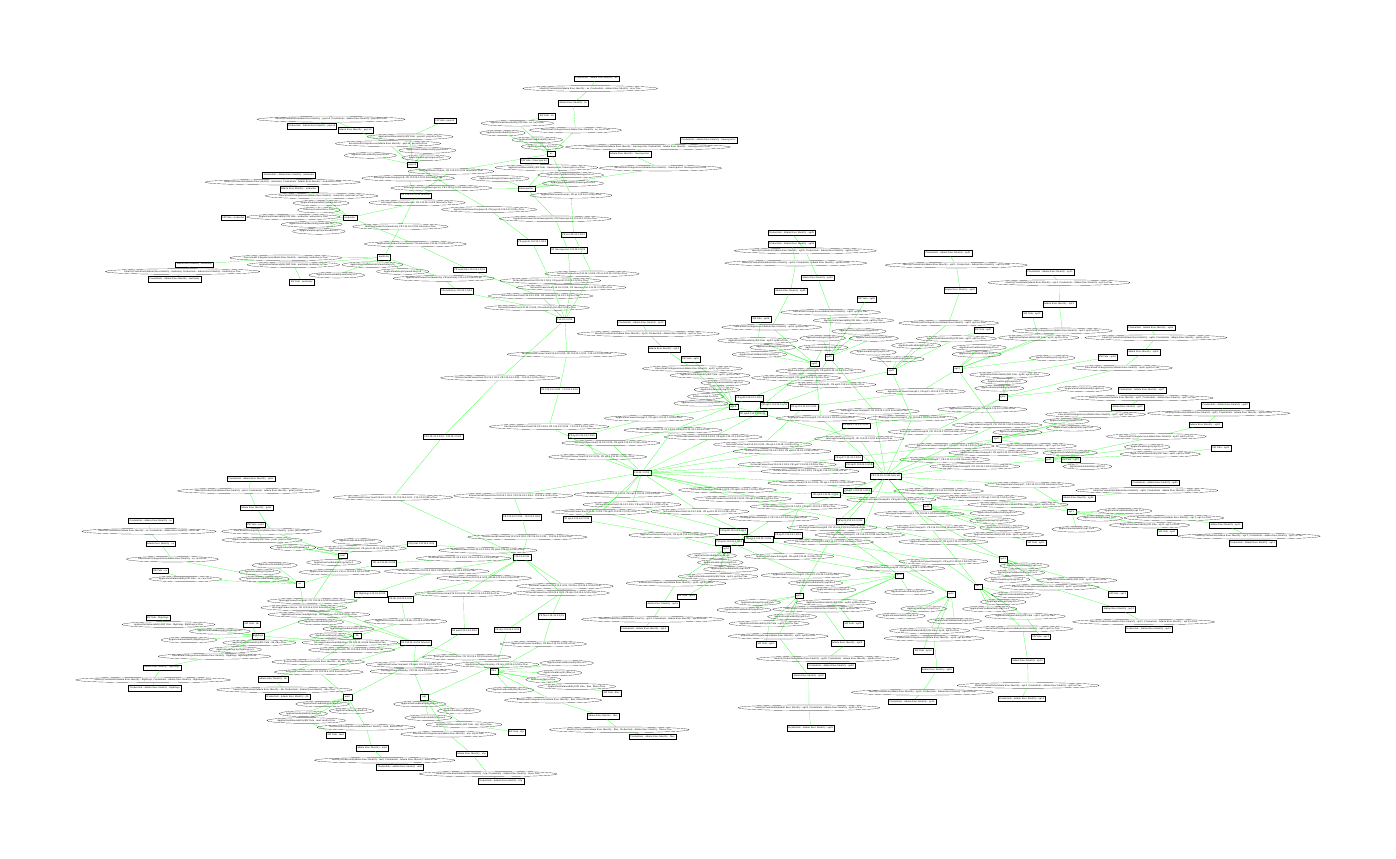}
    \caption{Bipartite graph representation of an initial observed CRATE state. }\label{fig:crate_obs}
\end{figure*}

\end{document}